\documentclass[11pt]{article}

\usepackage[margin=2.5cm]{geometry}
\usepackage{amsfonts}
\usepackage{amsmath}
\usepackage{amssymb}
\usepackage{tikz-cd}
\usepackage{braket}
\DeclareMathAlphabet{\pazocal}{OMS}{zplm}{m}{n}
\usepackage{blkarray} 
\usepackage{booktabs}
\usepackage{bbm}
\usepackage{mathrsfs}
\usepackage{systeme, mathtools}
\usepackage{authblk}
\usepackage[hidelinks]{hyperref}

\begin{document}
\title{Spin-network states for the Bianchi I and IX cosmological models from quantum constrained symmetries}

\author[1,2]{Matteo Bruno \footnote{\texttt{bruno@cpt.univ-mrs.fr}}}
\author[3,2]{Giovanni Montani\footnote{\texttt{giovanni.montani@enea.it}}}
\author[2]{Edoardo Maria Panno \footnote{\texttt{panno.1965750@studenti.uniroma1.it}}}

\affil[1]{Aix-Marseille Univ, Universit\'e de Toulon, CNRS, CPT, Marseille, France} 
\affil[2]{Physics Department, Sapienza University of Rome, P.le A. Moro 5, 00185 Roma, Italy}
\affil[3]{ENEA, C.R. Frascati (Rome), Italy Via E.\ Fermi 45, 00044 Frascati (Roma), Italy} 
\date{}

\maketitle

\begin{abstract}
In this work, we implement at the quantum level the gauge-fixing conditions that relate the homogeneous \(SU(2)\) gauge theory of Ashtekar variables, describing the classical cosmological sector, to the usual minisuperspace formulation. We impose the so-called divergence constraint, which fixes the gauge to the homogeneous one and, at the classical level, recovers a finite-dimensional phase space, together with the diagonal constraint, which imposes the diagonality of the metric and has already been extensively studied in the literature. We construct the corresponding quantum operators for general cosmological models and provide a suitable regularization in terms of holonomies and fluxes. We then show that a special class of homogeneous spin-network states describing the Bianchi I and Bianchi IX models satisfies the quantum gauge-fixing constraints, using techniques developed in Reduced Quantum Loop Gravity. This provides a quantum-level link between the cosmological theory with full \(SU(2)\) gauge symmetry and standard Loop Quantum Cosmology, by reason of the effective Abelian structure of the selected states.
\end{abstract}

\section{Introduction}
One of the longstanding questions in theoretical physics is whether the implementation of symmetries commutes with the quantization procedure. An interesting and timely example of this problem is provided by the canonical quantization of cosmological models within the framework of Loop Quantum Gravity (LQG) \cite{Thiemann_2007,Rovelli_Vidotto_2014}.

The application of the techniques of LQG to cosmological scenarios leads to Loop Quantum Cosmology (LQC)\cite{Ashtekar_Singh_2011,Banerjee_Calcagni_Martín-Benito_2012}. However, LQC cannot be regarded as the cosmological sector of LQG, since it is based on the quantization of a symmetry-reduced phase space that cannot be recovered by simply restricting the full quantum theory \cite{Bojowald_2000b,Bojowald_2000a,Bojowald_2001,Bojowald_2002,Bojowald_2003,Bojowald_2020,Ashtekar_Bojowald_Lewandowski_2003,Brunnemann_Fleischhack_2012,Fleischhack_2018}.

Indeed, LQC implements the connection variables directly in minisuperspace, without retaining the local $SU(2)$ gauge symmetry of the full theory. An interesting discussion of how the residual global $SU(2)$ symmetry, still present before gauge fixing in the homogeneous setting, leads to the characteristic Abelian structure of LQC is presented in Ref.~\cite{Bruno_Montani_2023a,Bruno_Montani_2023b}. More recently, a classical construction of homogeneous Ashtekar--Barbero variables preserving the local $SU(2)$ gauge freedom has been developed in Ref.~\cite{Bruno_2025a,Bruno_2025b}. Within this framework, homogeneous quantum states are naturally identified with cylindrical functions of $SU(2)$ holonomies associated with suitable symmetric graphs.

A proposal to recover cosmological states directly within the full LQG framework is provided by the so-called Quantum-Reduced Loop Gravity (QRLG) \cite{Alesci_Cianfrani_2013a,Alesci_Cianfrani_2013b,Alesci_Cianfrani_Rovelli_2013}. In its original formulation, QRLG follows the standard minisuperspace strategy of considering diagonal connections and fluxes. This is achieved by constructing a class of cubical spin-network states with large spin labels satisfying a quantum condition implementing the diagonal gauge.

In Ref.~\cite{Cianfrani_Marchini_Montani_2012,Cianfrani_Montani_2012a,Cianfrani_Montani_2012b}, the role of the diagonal gauge condition was investigated in detail at both the classical and the quantum levels. Furthermore, the construction of homogeneous Ashtekar--Barbero variables presented in Ref.~\cite{Bruno_2025b} shows that, in addition to the diagonal gauge, a divergence-free gauge condition is required to recover the standard homogeneous sector, i.e. the one underlying the usual formulation of LQC.

In the present work, we address the problem of constructing quantum states of Loop Quantum Gravity satisfying simultaneously the diagonal and divergence-free gauge conditions.

Our analysis focuses on homogeneous cosmological models, in particular the Bianchi I and Bianchi IX geometries. At the classical level, the simultaneous imposition of the diagonal and divergence-free conditions reproduces the reduced phase space of Loop Quantum Cosmology. It is therefore natural to investigate whether this property survives at the quantum level. We show that the simultaneous implementation of the corresponding quantum gauge-fixing conditions selects the same class of states introduced in Quantum-Reduced Loop Gravity.

This result provides a significant step towards answering the question of which additional conditions must be imposed on Loop Quantum Gravity states in order to recover a cosmological sector characterized by almost-periodic functions and an effective Abelian structure. Indeed, both in Loop Quantum Cosmology and in Quantum-Reduced Loop Gravity, the relevant $SU(2)$ representations reduce to those of three commuting $U(1)$ groups. In the present work, we show that this reduction naturally emerges as the consequence of implementing the appropriate quantum gauge-fixing conditions.

The paper is structured as follows. \hfill \break
In Sec.~\ref{sec:Review LQG}, we review the mathematical aspects of Loop Quantum Gravity.
We discuss the construction of quantum states, the fundamental operators associated with the canonical variables of the theory, and their explicit action on these states. 
Furthermore, we outline the implementation of the Gauss constraint, which plays an important role in the gauge-fixing procedure developed below. \hfill \break
In Sec.~\ref{sec:Review Bianchi I}, we present the loop quantization procedure for the diagonal Bianchi I model. 
Furthermore, we discuss the introduction of two gauge-fixing conditions—namely the diagonal constraint and the divergence constraint—that classically select the diagonal and homogeneous gauge upon which LQC is constructed. \hfill \break
In Sec.~\ref{sec: classical gauge-fixing}, we analyze the constraint algebra by computing the Poisson brackets between the gauge-fixing conditions and the canonical constraints of the theory. 
We establish the consistency of these conditions with the underlying constraint structure by evaluating the corresponding Faddeev--Popov matrix. \hfill \break
In Sec.~\ref{sec: quantum gauge operators}, we construct well-defined quantum operators associated with the classical gauge-fixing conditions by using the holonomy-flux representation.\hfill \break
In Sec.~\ref{sec: cubical states}, starting from the definition of homogeneous graphs, we construct cubical spin-network states in analogy with Quantum-Reduced Loop Gravity and verify that they satisfy the diagonal operator. \hfill \break
In Sec.~\ref{sec: res divergence constraint}, we evaluate the action of the divergence operator on the cubic states, showing that these states are also annihilated by the divergence constraint. \hfill \break
In Sec.~\ref{sec: Bianchi IX}, we review the loop quantization of the Bianchi IX model. Furthermore, we construct cubic states in analogy with the Bianchi I case and verify that, in this model as well, these states satisfy both quantum gauge-fixing conditions.

\section{Loop Quantum Gravity \label{sec:Review LQG}}
This section briefly reviews the foundations of Loop Quantum Gravity, establishes the notation, and outlines the core classical variables.
Furthermore, it describes the kinematic quantum states and the action of the fundamental quantum operators. \hfill \break
The fundamental classical variables of LQG are holonomies $h_e[A]$ along a path $e$, and fluxes $E_f[S]$ across a surface $S$. These represent a smeared version of the standard Ashtekar-Barbero variables $(A^i_a, E^b_j)$ \cite{Ashtekar_1986,Ashtekar_1987,Barbero-G._1995,Immirzi_1997} and are defined as:
\begin{equation}
    h_{e}[A] = \pazocal{P} \exp\left(- \int_0^1 ds \, \dot{e}^a A^i_a \tau_i \right),
\end{equation}
\begin{equation}
    E_f[S] = \int_S f^i E^{a}_i n_a \, d^2 u,
\end{equation}
where $s$ parametrizes the path $e$, $(u^1, u^2)$ denotes a coordinate system on the surface $S$, and $n_a$ is its normal one-form. The matrices $\tau_i$ are the anti-Hermitian generators of the $SU(2)$ gauge group, normalized according to $\operatorname{tr}(\tau_i \tau_j) = -\frac{1}{2}\delta_{ij}$, while $f: \Sigma \to \mathfrak{su}(2)$ is a smooth $\mathfrak{su}(2)$-valued function on the three-dimensional manifold $\Sigma$.
The kinematical Hilbert space of Loop Quantum Gravity, denoted as $\mathcal{H}_{\text{kin}}$, is defined as the Cauchy completion of the space of cylindrical functions with respect to the Ashtekar-Lewandowski measure \cite{Ashtekar_Lewandowski_1995}. 
Given a graph $\Gamma$, defined as a collection of $N$ edges $\{e_1, \dots, e_N\}$, a functional $\Psi[A]$ is said to be cylindrical with respect to $\Gamma$ if it depends on the connection $A$ exclusively through the holonomies along its edges:
\begin{equation}
        \Psi_{\Gamma}[A] = \psi(h_{e_1}[A], \dotsc, h_{e_N}[A]),
\end{equation}
where $\psi$ is a smooth function on the Lie group $SU(2)^N$.
By virtue of the Peter--Weyl theorem, an orthonormal basis for the space of cylindrical functions can be systematically constructed using the irreducible representation matrices of $SU(2)$, denoted as $D^{(j_e)}_{m_en_e}(h_e)$. Consequently, the functions
\begin{equation}
    f^{\Gamma}_{\{j_e \}, \{ m_e \}, \{ n_e \}}(h_{e_1}, \dotsc, h_{e_N}) = \prod_{e \in \Gamma} \sqrt{d_{j_e}} \, D^{(j_e)}_{m_en_e}(h_e)
    \label{basis for cylindrical function for a graph}
\end{equation}
span the space of cylindrical functions associated with the graph $\Gamma$, where $d_{j_e} = 2j_e + 1$ is the dimension of the representation, and the quantum numbers $j_e$, $m_e$, and $n_e$ range over all their allowed values.
The elementary operators in Loop Quantum Gravity are the holonomy and flux operators. The holonomy operator, associated with an edge $e$, acts on cylindrical functions via multiplication:
\begin{equation}
    \hat{D}^{(j)}_{mn} (h_e) \, \Psi_\Gamma[A] = D^{(j)}_{mn} (h_e) \, \psi(h_{e_1}, \dotsc ,h_{e_N}).
\end{equation}
The explicit result of this action depends on the topological relation between the edge $e$ and the graph $\Gamma$. If $e$ is not contained in $\Gamma$, the operator adds a new edge to the graph, yielding a cylindrical function defined on the extended graph $\Gamma' = \Gamma \cup e$. Conversely, if $\Gamma$ already contains $e$, we can decompose the cylindrical function into basis states—as in Eq.~\eqref{basis for cylindrical function for a graph}. In this case, only the factor associated with $e_i = e$ is affected by the multiplication:
\begin{equation}
    D^{(j)}_{mn} (h_e) \, D^{(j_{e_i})}_{m_{e_i}n_{e_i}} (h_{e_i}) \quad \text{for } e_i = e.
\end{equation}
This product can be explicitly evaluated using $SU(2)$ recoupling theory, which, for the simplest case involving two representations considered here, reduces to the application of Clebsch--Gordan coefficients.
To simplify the description of the flux operator, we introduce the auxiliary operator $\hat{J}_i^{(v,e)}$. Each of these operators carries an $SU(2)$ vector index $i$ and is labeled by a point $v$ and an edge $e$, where $v$ is either the source or the target point of $e$. The action of $\hat{J}_i^{(v,e)}$ on a cylindrical function based on $\Gamma$ is defined as:
\begin{equation}
    \hat{J}_i^{(v,e)} \, \Psi_{\Gamma}[A] = 
    \begin{cases}
        & i  \left.\frac{d}{d\varepsilon}\right|_{\varepsilon=0}\, \psi(h_{e_1}, \dotsc, h_{e_k}\, e^{\varepsilon \tau_i}, \dotsc , h_{e_N}) \quad \text{if } e=e_k \text{ and } e \text{ begins at } v \\
        & -i  \left.\frac{d}{d\varepsilon}\right|_{\varepsilon=0}\, \psi(h_{e_1}, \dotsc, e^{\varepsilon \tau_i} \,h_{e_k}, \dotsc ,h_{e_N}) \quad \text{if } e=e_k \text{ and } e \text{ ends at } v
    \end{cases}   
\end{equation}
If $e$ is not an edge of $\Gamma$ or if $v$ is not a node of the graph, the result is set to zero: $\hat{J}_i^{(v,e)} \, \Psi_{\Gamma}[A] = 0$. 
It is immediate to see that the action of $\hat{J}_i^{(v,e)}$ on a holonomy is given by:
\begin{equation}
    \hat{J}_i^{(v,e)} \, D^{(j)}(h_e) = 
    \begin{cases}
        & i \, D^{(j)}(h_e) \, \tau_i^{(j)} \\
        & -i \,  \tau_i^{(j)} \, D^{(j)}(h_e)
    \end{cases}   
\end{equation}
where $\tau_i^{(j)}$ are the generators of $SU(2)$ in the spin-$j$ representation. 
The flux operator can be expressed in terms of the operator $\hat{J}_i^{(v,e)}$ as 
\begin{equation}
    \hat{E}_i[S]\, \Psi_\Gamma[A] = \frac{\kappa \beta}{2} \sum_{x \in S} \sum_{\text{$e$ at $x$}} \nu(S, e) \, \hat{J}^{(x,e)}_i \, \Psi_\Gamma[A],
\end{equation}
where $\beta$ is the Barbero--Immirzi parameter, $\kappa$ is a constant defined as $\kappa = 8 \pi G$, and the geometric factor $\nu(S, e)$ is defined as:
\begin{equation}
\nu(S, e) = 
\begin{cases} 
+1 & \text{if } e \text{ lies above } S \\
0 & \text{if } e \text{ intersects } S \text{ tangentially or not at all} \\
-1 & \text{if } e \text{ lies below } S 
\end{cases}
\end{equation}
Here, ``above" and ``below" are defined relative to the direction of the normal vector of the surface.
The final operator required for our analysis is the Gauss constraint operator. Starting from the classical smeared expression of the Gauss constraint,
\begin{equation}
    G(\Lambda) = \frac{1}{\kappa \beta} \int_\Sigma d^3x \, \Lambda^i \left(\partial_aE^a_i + \epsilon_{ijk} \, A^j_a \, E^a_k \right),
\end{equation}
where $\Lambda^i(x)$ is a $\mathfrak{su}(2)$-valued smearing test function, one obtains a quantum operator whose action on cylindrical functions is given by:
\begin{equation}
    \hat{G}(\Lambda) \, \Psi_\Gamma[A] = \sum_{v \in \Gamma} \Lambda^i(v) \sum_{e \text{ at } v} \hat{J}_i^{(v, \, e)} \, \Psi_\Gamma[A].
\end{equation}
Here, the first sum runs over all vertices $v$ of the graph $\Gamma$, while the second regularizes the action by summing over all edges $e$ incident to the vertex $v$ (either as a source or a target).

\section{Loop quantization of the Bianchi I model \label{sec:Review Bianchi I}}
The Bianchi I model represents the simplest cosmological framework describing a homogeneous but anisotropic universe. 
This model can be quantized in the loop framework, as originally demonstrated by Ashtekar and Wilson-Ewing in \cite{Ashtekar_Wilson-Ewing_2009}. The geometry is described by the diagonal metric:
\begin{equation}
    ds^2 = -(N \, dt)^2 + (a_1 \, dx^1)^2 + (a_2 \, dx^2)^2 + (a_3 \, dx^3)^2
\end{equation}
where $a_i$ denote the directional scale factors and $N$ is the lapse function.
By exploiting the symmetries of this model, the Ashtekar variables reduce to:
\begin{equation}
    A^i_a = \frac{c^{(i)}}{L_{(I)}} \, \delta^i_I \, \theta^I_a, \quad
    E^a_i = \frac{\det(\theta)}{V_0} \, L_{(I)} \, p^{(i)} \, \delta^I_i \xi_I^a, 
\end{equation}
where $\xi^a_I$ and  $\theta^J_b$ represent the set of fiducial vectors and one-forms, respectively. The fiducial length $L_I$ characterizes the extension of an edge along the $I$-th direction with respect to the fiducial metric, and $V_0$ denotes the corresponding fiducial volume.
The symmetry-reduced parameters $c^i$ and $p_j$ completely characterize the connection and the triad fields. Their non-vanishing Poisson brackets read:
\begin{equation}
    \left\{c_i, p^j \right\} = \frac{\kappa \beta}{3} \delta^j_i.
\end{equation}
Furthermore, these variables are directly related to the physical scale factors and their time derivatives:
\begin{equation}
\begin{split}
    &\tilde{c}_1 = \frac{\beta}{N} \, L_1 \, \dot{a}_1, \quad
    \tilde{p}_1 = \text{sgn}(a_1) \, L_2 \, L_3 \, |a_2 \, a_3|, \\
    \label{chap. 3.3, momenta of Bianchi I}
    &\tilde{c}_2 = \frac{\beta}{N} \, L_2 \, \dot{a}_2, \quad 
    \tilde{p}_2 = \text{sgn}(a_2) \, L_1 \, L_3 \, |a_1 \, a_3|,\\
    &\tilde{c}_3 = \frac{\beta}{N} \, L_3 \, \dot{a}_3, \quad
    \tilde{p}_3 = \text{sgn}(a_3) \, L_1 \, L_2 \, |a_1 \, a_2| .\\
\end{split}
\end{equation}
At the quantum level, the kinematic basis states are of the form $\ket{p_1, p_2, p_3}$, which constitute a complete set of eigenstates of the quantum geometry operators. The action of the fundamental operators on these basis states is given by:
\begin{equation}
    \begin{split}
        & \hat{h}_{e_1}[A] \ket{p_1, p_2, p_3} = \ket{p_1 - \kappa \beta \mu_1, p_2, p_3}, \\ 
        & \hat{p}_1 \ket{p_1, p_2, p_3} = p_1 \ket{p_1, p_2, p_3},
    \end{split}
\end{equation}
with analogous relations holding for the remaining spatial directions. Here, $\mu_I$ is a real dimensionless parameter such that $\mu_I L_I$ represents the physical length of the edge $e_I$.

Since the formulation of Loop Quantum Cosmology, the formal link between LQC and the full Loop Quantum Gravity framework has remained elusive. In standard cosmological scenarios, once the full Ashtekar variables are replaced with their respective minisuperspace counterparts, both the Gauss and the spatial diffeomorphism constraints vanish identically. This feature is characteristic of LQC and implies the loss of a significant portion of the gauge symmetry inherent to the full theory. This restriction stems directly from the classical reduction of the phase space variables. 
Typically, the triads are chosen to be strictly parallel to the fiducial directions; however, this represents an arbitrary restriction. 
In order to justify this choice, a specific relation, named diagonal constraint, was introduced in \cite{Cianfrani_Montani_2012a,Cianfrani_Montani_2012b}. This constraint is uniquely satisfied by diagonal triads and must be implemented on generic triad fields to restrict the framework to the standard configurations adopted in LQC, which do not allow for internal spatial rotations. Consequently, this procedure seeks to identify the conventional choice of variables in LQC with a classical gauge-fixing procedure.
Moreover, while a homogeneous cosmological model is fundamentally characterized by homogeneous variables, the internal rotational invariance of the theory allows for both global and local arbitrary $SU(2)$ rotations of the triad fields. If the rotation matrix is permitted to be space-dependent, this gauge transformation risks generating an inhomogeneous class of variables.
To circumvent this issue, a well-defined class of generalized homogeneous variables has been proposed in the literature \cite{Bruno_2025b}. 
By going beyond the conventional minisuperspace framework, this mathematical formulation allows the gauge transformations to possess local spatial degrees of freedom while preserving the homogeneity of the variables. From a geometrical perspective, the complete constraint algebra of Loop Quantum Gravity is recovered, and the constraints take precisely the same form as in the original Ashtekar formulation. Consequently, a quantum theory closely resembling full LQG can be formulated by using these variables. Furthermore, it has been demonstrated that this framework can be mapped onto the original formulation of LQC for homogeneous models proposed by M.~Bojowald \cite{Bojowald_2013} via a local $SU(2)$ rotation of the triad fields.

\section{Gauge fixing on the classical phase space \label{sec: classical gauge-fixing}}
The main objective is to treat the constraints previously introduced in the literature as gauge-fixing conditions that restrict the full classical phase space of the Ashtekar formulation of General Relativity to a diagonal and homogeneous gauge. 
This procedure aims to find the homogeneous cosmological sector of the theory, which is expected to reproduce the standard framework of Loop Quantum Cosmology. 
The starting point for this analysis is the investigation of the corresponding constraint algebra. \\
These specific gauge conditions consist of the diagonal constraint, introduced as in \cite{Cianfrani_Montani_2012a,Cianfrani_Montani_2012b} and scaled by a multiplicative factor,
\begin{equation}
    \chi_i = \frac{1}{\kappa \beta} \, \epsilon_{ijk} \, \theta^j_a \, E^a_k,
    \label{Diagonal constraint}
\end{equation}
and the divergence constraint, likewise formulated as in \cite{Bruno_2025b} and scaled by a multiplicative factor,
\begin{equation}
    \rho_i = \frac{1}{\kappa \beta} \, \partial_a E^a_i.
    \label{Divergence constraint}
\end{equation}
By considering the smeared expressions of these constraints, $\pazocal{X}(\Omega)$ for the diagonal constraint and $\pazocal{R}(\Delta)$ for the divergence one, and utilizing the fundamental canonical Poisson brackets between the Ashtekar connection and the densitized triad,
\begin{equation}
    \left\{A^i_a(x), E^b_j(y) \right\} = \kappa \beta \, \delta^i_j \, \delta_a^b \, \delta^{(3)}(x,y),
\end{equation}
one can evaluate the Poisson brackets between the standard constraints of the theory and the newly imposed gauge-fixing conditions. 
Firstly, we focus on the divergence constraint.
Its Poisson brackets with the fundamental canonical variables are given by:
\begin{subequations}
    \begin{equation}
        \left\{ \pazocal{R}(\Delta), E^a_i(x) \right\} = 0,
        \label{divergence-connection bracket}
    \end{equation}
    \begin{equation}
        \left\{ A^i_a(x), \pazocal{R}(\Delta) \right\} 
        = - \partial_a\Delta^i(x).
        \label{divergence-triad bracket}
    \end{equation}
\end{subequations}
Eq.~\eqref{divergence-connection bracket} directly implies that the divergence constraint Poisson-commutes with itself:
\begin{equation}
    \left\{ \pazocal{R}(\Delta), \pazocal{R}(\tilde{\Delta}) \right\} = 0.
\end{equation}
Furthermore, the Poisson bracket with the Gauss constraint yields:
\begin{equation} 
        \left\{ G(\Lambda), \pazocal{R}(\Delta) \right\}
        = - \frac{1}{\kappa \beta} \int d^3x \, E^a_i \,  \epsilon_{ijk} \Lambda^j \, \partial_a\Delta^k \\
\end{equation}
Evidently, these two constraints do not commute. This is an expected result, as the choice of a homogeneous triad explicitly constitutes a gauge-fixing of the internal $SU(2)$ symmetry generated by the Gauss constraint. 
Regarding the diffeomorphism constraint, one gets:
\begin{equation}
    \left\{ D(\vec{U}), \pazocal{R}(\Delta) \right\}
    = \pazocal{R}(\pazocal{L}_U \Delta).
    \label{chap. 4.1, diff-divergence Poisson bracket result}
\end{equation}
The relation established in Eq.~\eqref{chap. 4.1, diff-divergence Poisson bracket result} indicates that, on shell, the Poisson bracket between the divergence and diffeomorphism constraints vanishes. Consequently, the divergence constraint does not act as a gauge-fixing condition for spatial diffeomorphism invariance; as expected, it exclusively breaks the internal $SU(2)$ symmetry for which it was originally introduced.
We now turn to the analysis of the diagonal constraint. Its Poisson brackets with the fundamental canonical variables are given by:
\begin{subequations}
    \begin{equation}
        \left\{ \pazocal{X}(\Omega), E^a_i(x) \right\} = 0,
        \label{diagonal-connectio bracket}
    \end{equation}
    \begin{equation}
        \left\{ A^i_a(x) , \pazocal{X}(\Omega) \right\} 
        = \epsilon_{ijk} \, \Omega^j(x) \, \theta^k_a(x).
        \label{diagonal-triad bracket}
    \end{equation}
\end{subequations}
As a consequence of Eq.~\eqref{diagonal-connectio bracket}, the diagonal constraint Poisson-commutes with itself:
\begin{equation}
    \left\{ \pazocal{X}(\Omega), \pazocal{X}(\tilde{\Omega}) \right\} = 0.
\end{equation}
Furthermore, the Poisson bracket with the Gauss constraint yields:
    \begin{equation}
        \left\{ G(\Lambda), \, \pazocal{X}(\Omega) \right\}= - \frac{1}{\kappa \beta} \int [ \Lambda^j(x) \, \Omega^s(x) \, E_k^a(x) \, \theta_a^k(x)- \Lambda^j(x) \, \Omega^k(x) \, E_k^a(x) \, \theta_a^j(x) ]\, d^3x\ .
    \end{equation} 
As expected, this bracket does not vanish, reflecting the fact that the diagonal gauge explicitly breaks the internal $SU(2)$ gauge symmetry.
Regarding the Poisson bracket with the diffeomorphism constraint, the computation is slightly more involved, yielding the final result:
    \begin{equation}
        \left\{ D(\vec{U}), \pazocal{X}(\Omega) \right\}=\pazocal{R}(\tilde{\Delta}) - \frac{1}{\kappa \beta} \int\epsilon_{ijk} \, E^{[a}_i \, U^{b]}  \left( \theta^k_a \, \partial_b\Omega^j - \frac{1}{2} \,\Omega^j \, C^k {}_{st} \, \theta^s_a \, \theta^t_b \right)\,d^3x
    \label{diff-diagonal Poisson bracket result}
\end{equation}
where we defined $\tilde{\Delta}^i := \epsilon_{ijk}\, U^b \, \Omega^j \, \theta^k_b$.
The relation established in Eq.~\eqref{diff-diagonal Poisson bracket result} is fully consistent with our expectations. The gauge-fixing condition restricting the framework to diagonal triads explicitly breaks spatial diffeomorphism invariance; consequently, the Poisson bracket between these two conditions is not expected to vanish on-shell.
Finally, we observe that the Poisson bracket between the two gauge-fixing conditions vanishes identically:
\begin{equation}
    \left\{ \pazocal{R}(\Delta), \pazocal{X}(\Omega) \right\} = 0.
\end{equation}
This crucial feature guarantees that the order in which the two gauge-fixing conditions are imposed is irrelevant.
To validate our gauge-fixing procedure, we compute the determinant of the matrix of the Poisson brackets between the first-class constraints of the theory and the gauge-fixing conditions.
Once the constraint hypersurface is selected and the gauge-fixing conditions are imposed, the only non vanishing parts of the matrix are the off diagonal blocks.
The determinant can be computed by considering only those, leading to:
    \begin{equation}
    \det \begin{pmatrix}
\{ G(\Lambda), \, \pazocal{X}(\Omega) \} & \{ G(\Lambda), \, \pazocal{R}(\Delta) \} \\
\{ D(\vec{U}), \, \pazocal{X}(\Omega) \} & 0 
\end{pmatrix}
\not\approx 0.
\label{determinant condition}
\end{equation}
Starting from the two constraints that implement the selection of homogeneous and diagonal phase space variables, we computed their Poisson brackets with the first-class constraints of the theory. 
The results summarized in Eq.~\eqref{determinant condition} guarantee that we are implementing a consistent set of gauge-fixing conditions for the $SU(2)$ and diffeomorphism symmetries \cite{Henneaux:1992ig}. 
The subsequent step involves the quantum implementation of this set of conditions on the states of the full theory. 
This is necessary to verify whether the resulting solutions exhibit a formal relation with the standard quantum states of Loop Quantum Cosmology.

\section{Quantum gauge fixing operators \label{sec: quantum gauge operators}}

In this section, we present the procedure to construct the quantum operators associated with the classical gauge-fixing conditions. Following the standard quantization prescription of Loop Quantum Gravity, these constraints are regularized and expressed in terms of holonomy and flux operators.
Even though the quantum states are discussed within the specific framework of the Bianchi I model, we present a construction of the quantum operators that remains entirely independent of the chosen cosmological model.
Regarding the diagonal constraint, we follow the procedure proposed in \cite{Alesci_Cianfrani_2013a}; specifically, we smear the fluxes along specific surfaces $S_I$ that are ``normal" to the direction $\xi_I$ and then consider the square of the constraint. 
The existence of a global coordinate system parallel to the directions identified by the left-invariant vector field $\xi_I$ is possible only in Bianchi I models. 
However, we can construct local coordinate systems using the exponential map and Riemann normal coordinates \cite{doCarmo1992,Carroll:2004st}.
It is well known that, given a point $p$ in a neighborhood $U$ of the three-dimensional spatial manifold $\sigma$ and a vector $v$ in the tangent space $T_p\sigma$, there exists a map $\gamma(s, p, v)$ which is the unique geodesic that passes through $p$ at $s=0$ with velocity $v$. 
Moreover, this map satisfies the property $\gamma(s, p, av) = \gamma(as, p, v)$.
The exponential map is defined as:
\begin{equation}
    \exp(p, v) = \gamma(1, p, v) = \gamma(|v|, p, \hat{v}).
\end{equation}
It is worth noting that this map differs from the standard exponential map of the Lie group. Indeed, the two definitions coincide exclusively when the Lie group is compact.
We now construct a set of local inertial coordinates by using the exponential map. The fiducial metric, expressed in terms of the left-invariant forms, reads:
\begin{equation}
    q_0 = \delta_{IJ} \, \theta^I \otimes \theta^J.
\end{equation}
We seek a coordinate system $\tilde{x}^I$ for which the basis vectors $\xi_I$ coincide with the coordinate basis, i.e., $\xi_I = \tilde{\partial}_I$. The exponential map achieves this condition automatically. Specifically, for any point $b$ within a sufficiently small neighborhood of $p$, there exists a unique geodesic path connecting $p$ to $b$, characterized by a unique parameterization $s$ such that $\gamma(s=0, p, v) = p$ and $\gamma(s=1, p, v) = b$. At the point $p$, the tangent vector $v$ to this geodesic can be decomposed as $v = v^I \xi_I$. We then define the sought-after coordinates $\tilde{x}^I$ of the point $b$ directly through these components: $\tilde{x}^I(b) = v^I$. Coordinates constructed in this manner define the Riemann normal coordinate system centered at $p$.
At this point, we can formally define the smearing of the flux on a surface $S_K$ normal to the direction $\xi_K$. 
Using our local coordinate system aligned with the fiducial directions, we consider a surface $S_K$, which is the set of points that satisfy $\tilde{x}^K = 0$, with volume form $d\tilde{x}^I \wedge d\tilde{x}^J$ and with normal vector parallel to $\xi_K$ (where $I \neq J \neq K$). 
Consequently, the flux along such a surface is defined as:
\begin{equation}
    E_i[S_K] = \frac{1}{2} \int_{S_K} \,  E^a_i \, \theta^K_a \, \frac{\epsilon_{KIJ}}{\det(\theta)} \, d\tilde{x}^I \wedge d\tilde{x}^J.
    \label{flux in riemann normal coordinates}
\end{equation}
If we choose the triads to be parallel to the fiducial directions, we can consider only lowercase Latin indices:
\begin{equation}
    E_i[S_k] = \frac{1}{2} \int_{S_k} \,  E^a_i \, \theta^k_a \, \frac{\epsilon_{kst}}{\det(\theta)} \, d\tilde{x}^s \wedge d\tilde{x}^t.
\end{equation}
By inserting the standard form of the triad in Loop Quantum Cosmology, i.e., $E^a_i = p_{(i)} \det(\theta) \xi^a_i$, into the previous integral, we find:
\begin{equation}
        E_i[S_k]  = \frac{1}{2} \int_{S_k} \, p_{(i)} \, \xi^a_i \, \theta^k_a \,  \epsilon_{kst} \,  d\tilde{x}^s \wedge d\tilde{x}^t = \frac{1}{2} \delta_{ik} \int_{S_k} \, p_{(i)} \, \epsilon_{kst} \, d\tilde{x}^s \wedge d\tilde{x}^t.
\end{equation} 
Therefore, the result is diagonal in the indices $i$ and $k$, and the diagonal constraint can be expressed as:
\begin{equation}
    \pazocal{X}_i = \epsilon_{ijk} \, E_j[S_k].
    \label{chap. 5.3, new smeared form of the diagonal constraint}
\end{equation}
Consider the square of the diagonal constraint, defined as:
\begin{equation}
    \pazocal{X}^2 
    = \sum_i \, \pazocal{X}_i \, \pazocal{X}_i \\  
    = \sum_{jk} \left( E_j[S_k] \, E_j[S_k] - E_j[S_k] \, E_k[S_j] \right).
    \label{chap.5.3, squared diagonal constraint}
\end{equation}
It is straightforward to promote the expression in Eq.~\eqref{chap.5.3, squared diagonal constraint} to a quantum operator:
\begin{equation}
    \hat{\pazocal{X}^2}
    = \sum_{jk} \left( \hat{E_j}[S_k] \, \hat{E_j}[S_k] - \hat{E_j}[S_k]\, \hat{E_k}[S_j] \right).
    \label{squared diagonal constraint}
\end{equation}

The regularization of the divergence constraint involves expressing the Ashtekar connection via holonomies defined on infinitesimal edges, alongside the smearing of the triad fields into fluxes over infinitesimal surfaces. By using the Gauss constraint $\pazocal{G}_i = (\kappa \beta)^{-1} \left(\partial_a E^a_i + \epsilon_{ijk}\, A^j_a \, E^a_k\right)$, we can rewrite $\rho_i$ as:
\begin{equation}
    \rho_i = \pazocal{G}_i - \frac{1}{\kappa \beta} \, \epsilon_{ijk} \, A^j_a \, E^a_k.
    \label{divergence constraint as Gauss minus NG}
\end{equation}
Upon introducing a suitable smearing function, the first term on the right-hand side can be straightforwardly identified with the Gauss constraint operator. Conversely, the second term must be appropriately regularized to obtain a well-defined quantum operator that accurately reproduces the classical constraint in the continuum limit of vanishing edge lengths.
To achieve this regularization, we consider the three-form $[A \wedge E]$, where $A = A^i_a \, \tau_i \, dx^a$ is the connection one-form and $E = E^a_i \, \epsilon_{abc} \, \tau_i \, dx^b \wedge dx^c$ denotes the flux two-form. Expanding the wedge product and the commutator, under a specific choice of orientation, yields the relation:
\begin{equation}
    [A \wedge E] = \epsilon_{ijk} \, A^i_a \, E^{a}_j \, \tau_k \, d^3x.
    \label{action of adj on algebra elements}
\end{equation}
From an algebraic perspective, the result in Eq.~\eqref{action of adj on algebra elements} represents the action of the adjoint representation of the $\mathfrak{su}(2)$ Lie algebra on $E$ mediated by $A$, namely $\text{ad}_A(E)$. Since the holonomy is an element of the $SU(2)$ gauge group rather than its algebra, it is natural to expect that the term in Eq.~\eqref{action of adj on algebra elements} can be recovered by taking the infinitesimal limit of the group adjoint representation, $\text{Ad}_{h_e}(E)$. 
By employing the previously introduced Riemann normal coordinates, we consider an infinitesimal edge oriented parallel to the direction $\xi_I$. We denote the holonomy along such an edge as $h_{c_I}[A]$, where $c_I$ represents the geodesic path $\gamma(s,p,\xi_I)$. For an infinitesimal domain $\left[- \frac{\varepsilon}{2} , \frac{\varepsilon}{2} \right]$ parameterizing $c_I$, the asymptotic expansion of the holonomy in the limit $\varepsilon \to 0$ reads:
\begin{equation}
    h_{c_I}[A] = 1 - \xi_I^a (p)A_a^i(p)\tau_i \ \varepsilon + O(\varepsilon^2).
\end{equation}
Regarding the flux, we introduce an infinitesimal surface $S_K$ normal to the direction $\xi_K$ at the point $p$. To formalize its infinitesimal nature, we define a square plaquette $S_K^{(\varepsilon)} := \left[- \frac{\varepsilon}{2}, \frac{\varepsilon}{2} \right] \times \left[- \frac{\varepsilon}{2}, \frac{\varepsilon}{2} \right]$. In Riemann normal coordinates, the flux through this surface, originally introduced in Eq.~\eqref{flux in riemann normal coordinates}, can be rewritten as:
\begin{equation}
    E_i \left[ S_K^{(\varepsilon)} \right] = \int_{S_K^{(\varepsilon)}} E^a_i \, \theta^K_a  \, \frac{n_K}{\det(\theta)} \, d^2\tilde{x}.
\end{equation}
By fixing a precise orientation, we can set the normal vector components to unity. The continuum limit $\varepsilon \to 0$ yields:
\begin{equation}
    E_i\left[ S_K^{(\varepsilon)} \right] = 
    \varepsilon^2 \, \frac{E^a_i(p) \, \theta^K_a(p)}{\det(\theta)(p)} + O(\varepsilon^3). 
\end{equation}
Now that both variables are well-defined, we can proceed to the explicit regularization of the divergence constraint. Recalling that the action of the adjoint representation of the $SU(2)$ gauge group is given by:
\begin{equation}
    \text{Ad}_{h_{c_I}[A]}(E[S_I]) =  h_{c_I}[A] \, E[S_I] \, h_{c_I}^{-1}[A],
\end{equation}
we can define a well-regularized smeared expression for the divergence constraint as follows:
\begin{equation}
       \pazocal{R}(\Delta) = G(\Delta) + \lim_{\varepsilon \to 0}\frac{\det(\theta)}{3 \kappa \beta \varepsilon^3} \sum_{I=1}^3 \left( h_{c_I}[A] \, E[S_I] \, h_{c_I}^{-1}[A] - E[S_I] \right).
    \label{smeared version of the divergence constraint}
\end{equation}  
By evaluating the limit explicitly, one straightforwardly recovers the non-gauge term originally introduced in Eq.~\eqref{divergence constraint as Gauss minus NG}.
It is worth noting that, in the limit $\varepsilon \to 0$, the first non-vanishing contribution appears at order $\varepsilon^3$. The leading-order term of order $\varepsilon^2$ automatically cancels out due to the subtraction of the flux term $E[S_I]$.
In order to define a quantum operator corresponding to the regularized version of the divergence constraint, a final step is required. The operators associated with the holonomy $h_{c_I}$ are expressed through the matrix elements $D^{(j)m} {}_{n}(h_{c_I})$. Since the adjoint representation of the $SU(2)$ group acts as the defining representation of $SO(3)$, the term $h_{c_I}[A] \, E[S_I] \, h_{c_I}^{-1}[A]$ can be interpreted as a spatial rotation of the flux operator in the internal space, mediated by a Wigner D-matrix in the $j=1$ representation.
Within the framework of Loop Quantum Gravity, the holonomy components are given by the Wigner D-matrix elements:
\begin{equation}
    D^{(j)m}_{\phantom{(j)m}n}(g) = \bra{j,m}\,g\, \ket{j,n},
\end{equation}
where $\ket{j,n}$ denotes the standard spherical basis in which the third generator $\tau_3=\tau_z$ is diagonal. To utilize the matrix $D^{(1)m}_{\phantom{(j)m}n}(h_e)$ to rotate the flux vector, the representation must be explicitly transformed into the Cartesian basis. This transformation is achieved through the change-of-basis matrix $U^i_{\phantom{i}m}$, defined as:
\begin{equation}
    U =
    \begin{pmatrix}
        \frac{i}{\sqrt{2}} & 0 & -\frac{i}{\sqrt{2}} \\
        \frac{1}{\sqrt{2}} & 0 & \frac{1}{\sqrt{2}} \\
        0 & i & 0
    \end{pmatrix},
\end{equation}
which diagonalizes the generator $\tau_3$. Consequently, the rotation of the flux field mediated by the holonomy is well-defined and reads:
\begin{equation}
    U^j{}_m \, D^{(1)m}_{\phantom{(j)m}n}(h_e) \,(U^\dagger)^n_{\phantom{n}i} \, E^i[S].
\end{equation}
The corresponding quantum operator is therefore given by:
\begin{equation}
    U^j{}_m \, \hat{D}^{(1)m}_{\phantom{(j)m}n}(h_e) \,(U^\dagger)^n_{\phantom{n}i} \, \hat{E}_i[S].
    \label{rotated flux}
\end{equation}
By using the expression for the rotated flux in Eq.~\eqref{rotated flux} and and specifying a particular, yet rather arbitrary, ordering of the operators, we can finally formulate the full quantum version of the divergence constraint:
\begin{equation}
    \begin{aligned}
            \hat{\pazocal{R}}(\Delta) 
            &= \hat{G}(\Delta) + \frac{\det(\theta)}{3\kappa \beta \varepsilon^3} \sum_{I=1}^{3} \Delta^i  \left(U^i{}_m \hat{D}^{(1)m}{}_n (h_{c_I}) (U^\dagger)^n{}_j \hat{E}_j [S_I] - \hat{E}_i [S_I] \right) \\
            &=: \hat{G}(\Delta) + \frac{\det(\theta)}{3\kappa  \beta \varepsilon^3} \sum_{i,I=1}^{3} \Delta^i \, \hat{R}^{i}_I.
    \end{aligned}
    \label{divergence constraint, final form}
\end{equation}
To ensure the correct contraction of the index structure, the factor $\Delta^i$ has been included inside the second term. This component acts as the local smearing function and is evaluated at the intersection point between the surface $S_I$ of the flux and the edge of the holonomy upon which the flux operator acts.

\section{Asymptotic cubical spin-network states \label{sec: cubical states}}
We now apply the defined operators to the quantum states of the theory. 
In the literature, the framework known as Quantum Reduced Loop Gravity (QRLG) proposes a specific class of states that solve the diagonal constraint given in Eq.~\eqref{squared diagonal constraint}.
A significant feature of these states is that, starting from the holonomy states of Loop Quantum Gravity, in a precise limit, they seem to reproduce the behavior of a product of three states, each characterized by a $U(1)$ symmetry \cite{Alesci_Cianfrani_2013a}.
This behavior is in analogy with the states of Loop Quantum Cosmology. 
We seek to construct an analogous state within the framework of the theory proposed in \cite{Bruno_2025b}, explicitly constructing homogeneous graphs. We focus our analysis on the Bianchi I model, where a global coordinate system parallel to the fiducial directions naturally exists. Specifically, Bianchi I can be equipped with a compact $\mathbb{T}^3$ topology, allowing one to construct a cubic state via three closed integral curves:
\begin{subequations}
    \begin{equation}
        \gamma_1(t) = (e^{it}, 1, 1),
    \end{equation}
    \begin{equation}
        \gamma_2(t) = (1, e^{it}, 1),
    \end{equation}
    \begin{equation}
        \gamma_3(t) = (1, 1, e^{it}).
    \end{equation}
\end{subequations}
The resulting graph $\Gamma = \gamma_1 \cup \gamma_2 \cup \gamma_3$ possesses a single vertex located at $(1, 1, 1)$, where all three curves intersect. A key property of $\Gamma$ is its homogeneity, resulting directly from its construction via integral curves of left-invariant vector fields; moreover, it constitutes a cubic graph in the sense of QRLG. This enables us to use the same tools developed in the QRLG framework, namely by considering edges with large spin quantum numbers ($j_e \gg 1$) such that each edge carries a representation matrix where both magnetic indices take either the maximal or minimal value ($j_e$ or $-j_e$) with respect to the basis aligned with the direction of the edge.
We denote the matrix elements of the Wigner D-matrices in the basis $\ket{j, m}_{ I}$ that diagonalizes the generator $\tau_{ I}$ as:
\begin{equation}
    D_{\mathrlap{ I} \phantom{(j)m} n}^{(j) m} (g) 
    = {}_{{ I}} \langle j,m \, | g \, | \, j,n\rangle_{{ I}}, \quad I= x,y,z.
\end{equation}
A generic basis state of the reduced Hilbert space has the form:
\begin{equation}
    \prod_e D_{\mathrlap{ I_e} \phantom{(j_e)\pm j_e} \pm j_e}^{(j_e) \pm j_e} (h_e),
\end{equation}
where the $+$ or $-$ sign corresponds to whether the edge $e$ is oriented identically or opposite to the coordinate axes. 
These states are not in the gauge-invariant Hilbert space, as they do not automatically satisfy the Gauss or the Diffeomorphism constraints; this reflects the imposition of the diagonal constraint. 
Furthermore, previous works indicates that these states do not possess any intertwiner; attempts to construct them yield only complex phases \cite{Bilski_Alesci_Cianfrani_Donà_Marcianò_2017}.
The absence of intertwiners is consistent with the formulation of Loop Quantum Cosmology.
If we keep track of the orientation of the graph, it is clear that:
\begin{equation}
    D_{\phantom{(j) j} j}^{(j)  j} (h^{-1})
    = D_{\phantom{(j) -j} -j}^{(j)  -j} (h).
\end{equation}
Therefore, we can focus on the the plus sign in the value of the magnetic numbers because the minus sign simply accounts to an inversion in the orientation of the edge of the holonomy.
A generic state constructed on the cubic graph takes the form:
\begin{equation}
    D_{\mathrlap{x} \phantom{(j_1)j_1}  j_1}^{(j_1) j_1} (h_1)\,
    D_{\mathrlap{y} \phantom{(j_2)j_2}  j_2}^{(j_2) j_2} (h_2)\,
    D_{\mathrlap{z} \phantom{(j_3)j_3}  j_3}^{(j_3) j_3} (h_3),
    \label{cubic state in Bianchi I}
\end{equation}
where $h_I := h_{c_I}$. In the three-torus model, we consider a single vertex where the three edges meet, with each edge entering and exiting this vertex.

To apply our quantum operators to the state in Eq.~\eqref{cubic state in Bianchi I}, we must first determine the action of the fundamental operators—fluxes and holonomies—on these reduced holonomies. 
We refer to the results established in \cite{Makinen_2020} for the necessary operational relations. \hfill \break
Regarding the flux operator, we require the action of the left and right invariant vector fields. Their action is identical and is given by the following relation:
\begin{equation}
    \hat{J}_i^{(v, e)} \, D_{\mathrlap{I} \phantom{(j)j}  j}^{(j) j} (h_{c_I}) =
    \begin{cases}
        & \pm j \, D_{\mathrlap{I} \phantom{(j)j}  j}^{(j) j} (h_{c_I}) \quad \text{if $i = {I}$}, \\
        & O(\sqrt{j}) \quad \text{if $i \neq {I}$}, \\
    \end{cases}
    \label{action of the flux on reduced holonomies}
\end{equation}
where the sign is $+$ if the edge $c_I$ originates from the vertex $v$, and $-$ if $c_I$ terminates at $v$. Consequently, in the large $j$ limit, the dominant contribution to the flux action occurs when the internal index matches the basis in which the Wigner D-matrix is computed, allowing us to neglect the remaining components. This result suggests that the non-commutative $SU(2)$ structure of the full theory effectively reduces, in the large $j$ limit, to a commutative $U(1)^3$ structure, echoing the framework of Loop Quantum Cosmology.
Next, we consider the holonomy operator. We distinguish two cases: one where the basis of the holonomy operator matches that of the reduced holonomy, and one where they differ. Furthermore, we shall focus exclusively on the operator $\hat{D}_{\mathrlap{z} \phantom{(1)j}  n}^{(1) m} (h_{c_I})$, which is the relevant component in the divergence constraint. \hfill \break
In the first case, we have:
\begin{equation}
    \hat{D}_{\mathrlap{z} \phantom{(1)m}  n}^{(1) m} (g) \, D_{\mathrlap{z} \phantom{(j)j}  j}^{(j) j} (g)
\end{equation} 
where both the operator and the state are along the same edge; otherwise, the action would reduce to a simple multiplication. The only significant results occur when $m=n$: 
\begin{subequations}
    \begin{equation}
        \hat{D}_{\mathrlap{z} \phantom{(1)1}  1}^{(1) 1} (g) \, D_{\mathrlap{z} \phantom{(j)j}  j}^{(j) j} (g) 
        = D_{\mathrlap{z} \phantom{(j+1)j+1}  j+1}^{(j+1) j+1} (g),
        \label{action of the holonomy on reduced holonomies 1}
    \end{equation}
    \begin{equation}
        \hat{D}_{\mathrlap{z} \phantom{(1)0}  0}^{(1) 0} (g) \, D_{\mathrlap{z} \phantom{(j)j}  j}^{(j) j} (g) 
        = D_{\mathrlap{z} \phantom{(j)j}  j}^{(j) j} (g) + O \left( \frac{1}{j} \right),
        \label{action of the holonomy on reduced holonomies 2}
    \end{equation}
    \begin{equation}
        \hat{D}_{\mathrlap{z} \phantom{(1)-1}  -1}^{(1) -1} (g) \, D_{\mathrlap{z} \phantom{(j)j}  j}^{(j) j} (g) 
        = D_{\mathrlap{z} \phantom{(j-1)j-1}  j-1}^{(j-1) j-1} (g) + O \left( \frac{1}{j} \right).
        \label{action of the holonomy on reduced holonomies 3}
    \end{equation}
\end{subequations}
All other combinations yield subdominant terms of order $O(1/j)$ or $O(1/\sqrt{j})$. 
In the second case, the bases differ:
\begin{equation}
   \hat{D}_{\mathrlap{z} \phantom{(1)m} n}^{(1) m} (g) D_{\mathrlap{I} \phantom{(j)j}  j}^{(j) j} (g), 
   \quad {I} = {x}, {y}.
\end{equation}
To evaluate this action, we must perform a change of basis to align the representation of the operator with that of the state. 
If we define $g_{{I}}$ as the rotation that maps the $z$-axis to the $I$-axis, we obtain the following transformation relation:
\begin{equation}
    \begin{split}
        \hat{D}_{\mathrlap{z} \phantom{(1)m} n}^{(1) m} (g) D_{\mathrlap{I} \phantom{(j)j}  j}^{(j) j} (g) =
        \sum_{m'} \,
        D_{\phantom{(1)m} m'}^{(1) m} (g_{I}) \,
        D_{\phantom{(1)m'} n}^{(1) m'} (g_{I}^{-1})
        \hat{D}_{\mathrlap{I} \phantom{(1)m'} m'}^{(1) m'} (g) \, 
        D_{\mathrlap{I} \phantom{(j)j}  j}^{(j) j} (g) \,
        + \text{off-diagonal terms}.
    \end{split}
    \label{action of the holonomy on reduced holonomies in different basis}
\end{equation}
The off-diagonal terms can be neglected because, once the holonomy operator is applied, the only relevant terms are those where the magnetic quantum numbers coincide.

Using these results, we can finally compute the action of the constraint operators on our state defined in Eq.~\eqref{cubic state in Bianchi I}. By construction, these states should satisfy the diagonal constraint; we can verify this explicitly:
\begin{equation}
    \hat{\pazocal{X}}^2 \,
    \left(D_{\mathrlap{x} \phantom{(j_1)j_1}  j_1}^{(j_1) j_1} (h_1)\,
    D_{\mathrlap{y} \phantom{(j_2)j_2}  j_2}^{(j_2) j_2} (h_2)\,
    D_{\mathrlap{z} \phantom{(j_3)j_3}  j_3}^{(j_3) j_3} (h_3) \right)
\end{equation}
Since the flux operator follows the Leibniz rule, we can focus on the three distinct terms. Consider the action on the holonomy along the first direction and recall the definition of the diagonal constraint in Eq.~\eqref{squared diagonal constraint}:
\begin{equation}
    \sum_{jk} \, \left( \hat{E}_j[S_k]  \, \hat{E}_j[S_k] -  \hat{E}_j[S_k]  \, \hat{E}_k[S_j]\right) D_{\mathrlap{x} \phantom{(j_1)j_1}  j_1}^{(j_1) j_1} (h_1).
\end{equation}
Due to the intersection factor $\nu(S_k, \, c_1)$, the action of the flux vanishes when $k \neq 1$, as the curve would be tangent to the surface. The only non-vanishing case occurs when the curve $c_1$ is normal to the surface $S_k$ at the intersection point, i.e., when $k=1$. Therefore, the expression reduces to:
\begin{equation}
    \sum_{j} \, \left( \hat{E}_j[S_1]  \, \hat{E}_j[S_1] -  \hat{E}_1[S_j]  \, \hat{E}_j[S_1]\right) D_{\mathrlap{x} \phantom{(j_1)j_1}  j_1}^{(j_1) j_1} (h_1).
\end{equation} 
Now, Eq.~\eqref{action of the flux on reduced holonomies} tells us that the only relevant term of this action is the one with $j=1$; hence we are left with:
\begin{equation}
    \left( \hat{E}_1[S_1]  \, \hat{E}_1[S_1] -  \hat{E}_1[S_1]  \, \hat{E}_1[S_1]\right) D_{\mathrlap{x} \phantom{(j_1)j_1}  j_1}^{(j_1) j_1} (h_1).
\end{equation}
It is clear that the result vanishes:
\begin{equation}
    \hat{\pazocal{X}}^2 \, D_{\mathrlap{x} \phantom{(j_1)j_1}  j_1}^{(j_1) j_1} (h_1) 
    = \left[(j_1  \kappa \beta)^2 - (j_1  \kappa \beta)^2 \right] D_{\mathrlap{x} \phantom{(j_1)j_1}  j_1}^{(j_1) j_1} (h_1) = 0.
    \label{risultato diagonale-cubi}
\end{equation}
To be precise, the action of the flux generates two terms because the curve intersects the surface twice: once at the starting point and once at the ending point. However, due to the intersection factor $\nu(S, e)$ in the flux action, it is straightforward to verify that the second term is identical to the one reported in Eq.~\eqref{risultato diagonale-cubi}. Exactly the same mechanism occurs for every edge; thus, we finally obtain:
\begin{equation}
    \hat{\pazocal{X}}^2 \,
    \left[D_{\mathrlap{x} \phantom{(j_1)j_1}  j_1}^{(j_1) j_1} (h_1)\,
    D_{\mathrlap{y} \phantom{(j_2)j_2}  j_2}^{(j_2) j_2} (h_2)\,
    D_{\mathrlap{z} \phantom{(j_3)j_3}  j_3}^{(j_3) j_3} (h_3) \right] = 0,
\end{equation}
as expected.

\section{Resolution of the divergence constraint \label{sec: res divergence constraint}}

Regarding the divergence operator in Eq.~\eqref{divergence constraint, final form}, the computation can be divided into two parts. The first deals with the Gauss constraint, while the second concerns the terms $\hat{R}^i_I$. \hfill \break
Recalling the action of the Gauss constraint and using Eq.~\eqref{action of the flux on reduced holonomies} we find:
\begin{equation}
    \hat{G}(\Lambda) \, D_{\mathrlap{x} \phantom{(j_1)j_1}  j_1}^{(j_1) j_1} (h_1)
    =\Lambda^1(v) \left[ -j_1 + j_1\right]D_{\mathrlap{x} \phantom{(j_1)j_1}  j_1}^{(j_1) j_1} (h_1) = 0.
\end{equation}
Therefore, the action on the full state vanishes:
\begin{equation}
    \hat{G}(\Lambda) \left[ 
    D_{\mathrlap{x} \phantom{(j_1)j_1}  j_1}^{(j_1) j_1} (h_1)\,
    D_{\mathrlap{y} \phantom{(j_2)j_2}  j_2}^{(j_2) j_2} (h_2)\,
    D_{\mathrlap{z} \phantom{(j_3)j_3}  j_3}^{(j_3) j_3} (h_3) \right] 
    = 0.
\end{equation}
To fully satisfy the divergence constraint, we must investigate the following relation:
\begin{equation}
    \sum_{I=1}^{3} \Delta^i \, \hat{R}^{i}_I \left[D_{\mathrlap{x} \phantom{(j_1)j_1}  j_1}^{(j_1) j_1} (h_1)\,
    D_{\mathrlap{y} \phantom{(j_2)j_2}  j_2}^{(j_2) j_2} (h_2)\,
    D_{\mathrlap{z} \phantom{(j_3)j_3}  j_3}^{(j_3) j_3} (h_3) \right].
\end{equation} 
Since the full state possesses only a single vertex, the term depending on the smearing function reduces to a multiplicative factor $\Delta^i(v)$ in front of each term in the sum. We consider the action of the three terms separately, keeping the $i$-index fixed, before making the sum explicit. Consequently, we focus on the following relation:
\begin{equation}
    \hat{R}_{i} \Bigl[ D_{\mathrlap{x} \phantom{(j_1)j_1} j_1}^{(j_1) j_1} (h_1) 
    D_{\mathrlap{y} \phantom{(j_2)j_2} j_2}^{(j_2) j_2} (h_2) 
    D_{\mathrlap{z} \phantom{(j_3)j_3} j_3}^{(j_3) j_3} (h_3) \Bigr]
\end{equation}
Again, we can focus separately on the action on the three different holonomies. Firstly, consider the $D_{\mathrlap{z} \phantom{(j_3)j_3} j_3}^{(j_3) j_3} (h_3)$ term.
By the same arguments already described and recalling the form of the divergence constrain in Eq.~\eqref{divergence constraint, final form}, we get:
\begin{equation}
    \begin{aligned}
        & \hat{R}_{i} \, D_{\mathrlap{z} \phantom{(j_3)j_3} j_3}^{(j_3) j_3} (h_3) = j_3 \frac{\kappa \beta}{2} \Bigl[ U^i{}_1 \, (U^\dagger)^1{}_3 \, D^{(j_3+1)j_3+1}_{\mathrlap{z} \phantom{(j_3+1)j_3+1} j_3+1} +  U^i{}_0 \, (U^\dagger)^0{}_3 \, D^{(j_3)j_3}_{\mathrlap{z} \phantom{(j_3)j_3} j_3} +\\
        & \qquad \qquad \qquad \qquad   + U^i{}_{-1} \, (U^\dagger)^{-1}{}_3 \, D^{(j_3-1)j_3-1}_{\mathrlap{z} \phantom{(j_3-1)j_3-1} j_3-1} - \delta_{i3} \, D^{(j_3)j_3}_{\mathrlap{z} \phantom{(j_3)j_3} j_3} \Bigr].
    \end{aligned}
\label{termine in z}
\end{equation} 
Here, we applied the action of the flux operator in Eq.~\eqref{action of the flux on reduced holonomies} and the full action of the holonomy from Eq.~\eqref{action of the holonomy on reduced holonomies 1}, Eq.~\eqref{action of the holonomy on reduced holonomies 2}, and Eq.~\eqref{action of the holonomy on reduced holonomies 3}
Now consider the $D_{\mathrlap{x} \phantom{(j_1)j_1} j_1}^{(j_1) j_1} (h_1)$-term. By applying the previously derived operational steps, we obtain a similar result:
\begin{equation}
    \begin{aligned}
        & \hat{R}_{i} \, D_{\mathrlap{x} \phantom{(j_1)j_1} j_1}^{(j_1) j_1} (h_1) = j_1 \frac{\kappa \beta}{2} \biggl\{ U^i{}_m \, (U^\dagger)^n{}_1 \\
        & \times \Bigl[ D^{(1)m}_{\phantom{(1)m} 1} (g_{x}) \, D^{(1) 1}_{\phantom{(1)1} n} (g_{x}^{-1}) \, D^{(j_1+1) j_1+1}_{\mathrlap{x} \phantom{(j_1+1)j_1+1} j_1+1} + D^{(1)m}_{\phantom{(1)m} 0} (g_{x}) \, D^{(1) 0}_{\phantom{(1)0} n} (g_{x}^{-1}) \, D^{(j_1) j_1}_{\mathrlap{x} \phantom{(j_1)j_1} j_1} \\
        & \qquad \qquad + D^{(1)m}_{\phantom{(1)m} -1} (g_{x}) \, D^{(1) -1}_{\phantom{(1)-1} n} (g_{x}^{-1}) \, D^{(j_1-1) j_1-1}_{\mathrlap{x} \phantom{(j_1-1)j_1-1} j_1-1} \Bigr] - \delta_{i1} \, D^{(j_1) j_1}_{\mathrlap{x} \phantom{(j_1)j_1} j_1} \biggr\}.\\
    \end{aligned}
\end{equation}
The main difference is the use of the basis rotation defined in Eq.~\eqref{action of the holonomy on reduced holonomies in different basis}, which leads to the presence of the rotation matrices $D_{\phantom{(1)}}^{(1)} (g_{x})$ and $D_{\phantom{(1)}}^{(1)} (g_{x}^{-1})$.
For the $D_{\mathrlap{y} \phantom{(j_2)j_2} j_2}^{(j_2) j_2} (h_2)$ term, we obtain a similar result:
\begin{equation}
    \begin{split}
        & \hat{R}_{i} \, D_{\mathrlap{y} \phantom{(j_2)j_2} j_2}^{(j_2) j_2} (h_2) 
        = j_2 \frac{\kappa \beta}{2} \biggl\{ U^i{}_m \, (U^\dagger)^n{}_2 \\
        & \times \Bigl[ D^{(1)m}_{\phantom{(1)m} 1} (g_{y}) \, D^{(1) 1}_{\phantom{(1)1} n} (g_{y}^{-1}) \, D^{(j_2+1) j_2+1}_{\mathrlap{y} \phantom{(j_2+1)j_2+1} j_2+1} + D^{(1)m}_{\phantom{(1)m} 0} (g_{y}) \, D^{(1) 0}_{\phantom{(1)0} n} (g_{y}^{-1}) \, D^{(j_2) j_2}_{\mathrlap{y} \phantom{(j_2)j_2} j_2} \\
        & \qquad \qquad  + D^{(1)m}_{\phantom{(1)m} -1} (g_{y}) \, D^{(1) -1}_{\phantom{(1)-1} n} (g_{y}^{-1}) \, D^{(j_2-1) j_2-1}_{\mathrlap{y} \phantom{(j_2-1)j_2-1} j_2-1} \Bigr] - \delta_{i2} \, D^{(j_2) j_2}_{\mathrlap{y} \phantom{(j_2)j_2} j_2} \biggr\}.
    \end{split}
\end{equation}
Now, notice that the terms of the following form:
\begin{equation}
    D^{(1)m}_{\phantom{(1)m} (n')} (g_{I}) \, D^{(1) (n')}_{\phantom{(1)(n')} n} (g_{I}^{-1}), \quad \text{with $n'$ fixed}
\end{equation}
correspond to the projector onto the state $\ket{1,n'}_{{I}}$, namely:
\begin{equation}
    D^{(1)m}_{\phantom{(1)m} (n')} (g_{I}) \, D^{(1) (n')}_{\phantom{(1)(n')} n} (g_{I}^{-1})  = {}^m(| 1, n' \rangle_{I} \, {}_{I} \langle n', 1 |)_n\eqqcolon (P^{({I})}_{n'})^m_{\phantom{m}n}. 
\end{equation}
This holds for both the $g_{x}$ and $g_{y}$ rotations, defining the projectors for the respective ${x}$- and ${y}$-bases:
\begin{subequations}
    \begin{equation}
        D^{(1)m}_{\phantom{(1)m} (n')} (g_{x}) \, D^{(1) (n')}_{\phantom{(1)(n')} n} (g_{x}^{-1})  = {}^m(| 1, n' \rangle_{x} \, {}_{x} \langle n', 1 |)_n \eqqcolon (P^{({x})}_{n'})^m_{\phantom{m}n},
    \end{equation}
    \begin{equation}
        D^{(1)m}_{\phantom{(1)m} (n')} (g_{y}) \, D^{(1) (n')}_{\phantom{(1)(n')} n} (g_{y}^{-1}) = {}^m(| 1, n' \rangle_{y} \, {}_{y} \langle n', 1 |)_n \eqqcolon (P^{({y})}_{n'})^m_{\phantom{m}n}.
    \end{equation}
\end{subequations}
Furthermore, every time these projectors appear, they are accompanied by a change of basis from the spherical to the Cartesian representation using the $U$ and $U^\dagger$ matrices. This allows us to define the projectors directly in the Cartesian basis through the following relations:
\begin{subequations}
    \begin{equation}
        U^i{}_m (P^{({x})}_{n'})^m_{\phantom{m}n} (U^\dagger)^n{}_j =: (U P^{({x})}_{n'} U^\dagger)^i{}_j,
    \end{equation}
    \begin{equation}
        U^i{}_m (P^{({y})}_{n'})^m_{\phantom{m}n} (U^\dagger)^n{}_j =: (U P^{({y})}_{n'} U^\dagger)^i{}_j,
    \end{equation}
    \begin{equation}
        U^i{}_{(n')} (U^\dagger)^{(n')}{}_j =: (U P^{({z})}_{n'} U^\dagger)^i{}_j,
    \end{equation}
\end{subequations}
where $n'$ is fixed.
By putting all the components together, we arrive at the final expression:
    \begin{equation}
    \begin{aligned}
        &\hat{R}_{i} \Bigl[ 
        D_{\mathrlap{x} \phantom{(j_1)j_1} j_1}^{(j_1) j_1} (h_1) 
        D_{\mathrlap{y} \phantom{(j_2)j_2} j_2}^{(j_2) j_2} (h_2) 
        D_{\mathrlap{z} \phantom{(j_3)j_3} j_3}^{(j_3) j_3} (h_3) 
        \Bigr] \\
        & \qquad =\frac{\kappa \beta}{2} \biggl\{ D^{(j_1) j_1}_{\mathrlap{x} \phantom{(j_1)j_1} j_1} \, D^{(j_2) j_2}_{\mathrlap{y} \phantom{(j_2)j_2} j_2} \, D^{(j_3) j_3}_{\mathrlap{z} \phantom{(j_3)j_3} j_3} \Bigl[ j_3 \left( (U P^{({z})}_0 U^\dagger)^i{}_3 - \delta_{i3} \right) \\
        & \qquad \qquad \qquad \qquad  + j_1 \left( (U P^{({x})}_0 U^\dagger)^i{}_1 - \delta_{i1} \right) 
        + j_2 \left( (U P^{({y})}_0 U^\dagger)^i{}_2 - \delta_{i2} \right) \Bigr] \\
        & \qquad \qquad \quad + D^{(j_1+1) j_1+1}_{\mathrlap{x} \phantom{(j_1+1)j_1+1} j_1+1} \, D^{(j_2) j_2}_{\mathrlap{y} \phantom{(j_2)j_2} j_2} \, D^{(j_3) j_3}_{\mathrlap{z} \phantom{(j_3)j_3} j_3} \Bigl( j_1 (U P^{({x})}_1 U^\dagger)^i{}_1 \Bigr) \\
        & \qquad \qquad \quad + D^{(j_1-1) j_1-1}_{\mathrlap{x} \phantom{(j_1-1)j_1-1} j_1-1} \, D^{(j_2) j_2}_{\mathrlap{y} \phantom{(j_2)j_2} j_2} \, D^{(j_3) j_3}_{\mathrlap{z} \phantom{(j_3)j_3} j_3} \Bigl( j_1 (U P^{({x})}_{-1} U^\dagger)^i{}_1 \Bigr) \\
        & \qquad \qquad \quad + D^{(j_1) j_1}_{\mathrlap{x} \phantom{(j_1)j_1} j_1} \, D^{(j_2+1) j_2+1}_{\mathrlap{y} \phantom{(j_2+1)j_2+1} j_2+1} \, D^{(j_3) j_3}_{\mathrlap{z} \phantom{(j_3)j_3} j_3} \Bigl( j_2 (U P^{({y})}_1 U^\dagger)^i{}_2 \Bigr) \\ 
        & \qquad \qquad \quad + D^{(j_1) j_1}_{\mathrlap{x} \phantom{(j_1)j_1} j_1} \, D^{(j_2-1) j_2-1}_{\mathrlap{y} \phantom{(j_2-1)j_2-1} j_2-1} \, D^{(j_3) j_3}_{\mathrlap{z} \phantom{(j_3)j_3} j_3} \Bigl( j_2 (U P^{({y})}_{-1} U^\dagger)^i{}_2 \Bigr) \\
        & \qquad \qquad \quad + D^{(j_1) j_1}_{\mathrlap{x} \phantom{(j_1)j_1} j_1} \, D^{(j_2) j_2}_{\mathrlap{y} \phantom{(j_2)j_2} j_2} \, D^{(j_3+1) j_3+1}_{\mathrlap{z} \phantom{(j_3+1)j_3+1} j_3+1} \Bigl( j_3 (U P^{({z})}_1 U^\dagger)^i{}_3 \Bigr) \\
        & \qquad \qquad \quad + D^{(j_1) j_1}_{\mathrlap{x} \phantom{(j_1)j_1} j_1} \, D^{(j_2) j_2}_{\mathrlap{y} \phantom{(j_2)j_2} j_2} \, D^{(j_3-1) j_3-1}_{\mathrlap{z} \phantom{(j_3-1)j_3-1} j_3-1} \Bigl( j_3 (U P^{({z})}_{-1} U^\dagger)^i{}_3 \Bigr) \biggr\}. \\
    \end{aligned}
    \label{final result of the non gauss part of div}
\end{equation}
Let us examine the explicit form of these projectors in the Cartesian basis:
\begin{equation*}
    U P^{({x})}_1 U^\dagger =
    \begin{pmatrix} 
    0 & 0 & 0 \\ 
    0 & \frac{1}{2} & -\frac{i}{2} \\ 
    0 & \frac{i}{2} & \frac{1}{2} 
    \end{pmatrix}, 
    \quad
    U P^{({x})}_0 U^\dagger =
    \begin{pmatrix} 
    1 & 0 & 0 \\ 
    0 & 0 & 0 \\ 
    0 & 0 & 0 
    \end{pmatrix}, 
    \quad
    U P^{({x})}_{-1} U^\dagger =
    \begin{pmatrix} 
    0 & 0 & 0 \\ 
    0 & \frac{1}{2} & \frac{i}{2} \\ 
    0 & -\frac{i}{2} & \frac{1}{2} 
    \end{pmatrix},
\end{equation*}
\begin{equation*}
    U P^{({y})}_1 U^\dagger =
    \begin{pmatrix} 
    \frac{1}{2} & 0 & -\frac{i}{2} \\ 
    0 & 0 & 0 \\ 
    \frac{i}{2} & 0 & \frac{1}{2} 
    \end{pmatrix}, 
    \quad
     U P^{({y})}_0 U^\dagger =
    \begin{pmatrix}
    0 & 0 & 0 \\ 
    0 & 1 & 0 \\ 
    0 & 0 & 0 
    \end{pmatrix}, 
    \quad U P^{({y})}_{-1} U^\dagger =
    \begin{pmatrix} 
    \frac{1}{2} & 0 & 
    \frac{i}{2} \\ 
    0 & 0 & 0 \\ 
    -\frac{i}{2} & 0 & \frac{1}{2} 
    \end{pmatrix}
\end{equation*}
\begin{equation*}
    U P^{({z})}_{1} U^\dagger =
    \begin{pmatrix} 
    \frac{1}{2} & \frac{i}{2} & 0 \\ 
    -\frac{i}{2} & \frac{1}{2} & 0 \\ 
    0 & 0 & 0 
    \end{pmatrix}, 
    \quad U P^{({z})}_{0} U^\dagger =
    \begin{pmatrix} 0 & 0 & 0 \\ 
    0 & 0 & 0 \\ 
    0 & 0 & 1 \end{pmatrix}, 
    \quad 
    U P^{({z})}_{-1} U^\dagger =
    \begin{pmatrix} 
    \frac{1}{2} & -\frac{i}{2} & 0 \\ 
    \frac{i}{2} & \frac{1}{2} & 0 \\ 
    0 & 0 & 0 
    \end{pmatrix}
\end{equation*}

By substituting these matrix values into Eq.~\eqref{final result of the non gauss part of div}, one can verify that the terms within the brackets cancel out for every value of $i$. Specifically, the terms associated with the $n'=0$ projectors vanish by direct subtraction, while the $n' = \pm 1$ ones vanish due to the specific form of the projectors. Consequently, we find:
\begin{equation}
    \begin{split}
        \hat{R}_{i} & \Bigl[ 
        D_{\mathrlap{x} \phantom{(j_1)j_1} j_1}^{(j_1) j_1} (h_1) 
        D_{\mathrlap{y} \phantom{(j_2)j_2} j_2}^{(j_2) j_2} (h_2) 
        D_{\mathrlap{z} \phantom{(j_3)j_3} j_3}^{(j_3) j_3} (h_3) 
        \Bigr] = 0.
    \end{split}
    \label{eq. vanishing of the regularized term of div constraint}
\end{equation}
Since both the Gauss term and the regularized non-Gauss term of the divergence constraint vanish independently, we have established the following relation:
\begin{equation}
    \hat{\pazocal{R}}(\Delta) \Bigl[ 
        D_{\mathrlap{x} \phantom{(j_1)j_1} j_1}^{(j_1) j_1} (h_1) 
        D_{\mathrlap{y} \phantom{(j_2)j_2} j_2}^{(j_2) j_2} (h_2) 
        D_{\mathrlap{z} \phantom{(j_3)j_3} j_3}^{(j_3) j_3} (h_3) 
        \Bigr] = 0.
\end{equation}

\section{The Bianchi IX model \label{sec: Bianchi IX}}

The Bianchi IX model is the most general homogeneous and anisotropic cosmological model invariant under three-dimensional spatial rotations. Being characterized by an $SO(3)$ symmetry, its left-invariant vector fields satisfy the standard $\mathfrak{su}(2)$ algebra, as do the fiducial vector fields:
\begin{equation}
    [\xi_I, \xi_J] = -\frac{2}{r_0} \, \epsilon^{K}_{\phantom{K}IJ} \, \xi_K.
\end{equation}
Hence, the structure constants are given by the completely antisymmetric tensor $\epsilon^{K}_{\phantom{K}IJ}$ scaled by the factor $2/r_0$, where $r_0$ denotes the radius of the $3$-sphere with respect to the fiducial metric. 
Due to the symmetry underlying this model, there exists an equivalence class of phase space variables related by an internal gauge rotation matrix. Consequently, we can select a specific pair of the form:
\begin{equation}
    A^i_a = \frac{c^{(i)}}{l_0} \, \delta^i_I \, \theta^I_a, \quad E^a_i = \frac{p_{(i)}}{l_0^2} \, \det(\theta) \, \delta^I_i \, \xi^a_I.
\end{equation}
Here, $l_0 = V_0^{1/3}$, where $V_0$ represents the fiducial volume. The physical information of the model is encoded within the gauge-invariant functions $c^i$ and $p_i$, which satisfy the canonical Poisson brackets:
\begin{equation}
    \left\{c_i, p^j \right\} = \kappa\beta \, \delta^j_i.
\end{equation}
These connection variables are directly related to the metric formalism; specifically, $p_{i} = \text{sgn}(a_i)|a_ja_k| l_0^2$ with $i \neq j \neq k$, while the $c^i$ are determined through the explicit computation of the spin connection $\Gamma$ and the extrinsic curvature $K$.
The closed FRW model is recovered by setting $a_1 = a_2 = a_3 = a$ in the Bianchi IX phase space.
As in the Bianchi I model, at the quantum level the kinematic basis states are of the form $\ket{p_1, p_2, p_3}$, which constitute a complete set of eigenstates of the quantum geometry operators \cite{Wilson-Ewing_2010}. The action of the fundamental operators on these basis states is given by:
\begin{equation}
    \begin{split}
        & \hat{h}_{e_1}[A] \ket{p_1, p_2, p_3} = \ket{p_1 - \kappa \beta \mu_1, p_2, p_3}, \\ 
        & \hat{p}_1 \ket{p_1, p_2, p_3} = p_1 \ket{p_1, p_2, p_3},
    \end{split}
\end{equation}
with analogous relations holding for the remaining spatial directions. 

At the quantum level, the Bianchi IX model can be equipped with a compact $\mathbb{S}^3$ topology. A generic graph for this model is composed of three holonomies, each constructed along integral curves parallel to the fiducial directions of the sphere. Identifying the 3-sphere with the group $SU(2)$, these integral curves are generated via the Lie group exponential map, which, due to the compactness of the group, coincides with the Riemannian exponential map and yields the following expressions:
\begin{subequations}
    \begin{equation}
        \gamma_1(t) = \begin{pmatrix}
            \cos(\frac{t}{2}) \quad &-i\sin(\frac{t}{2}) \\
            -i\sin(\frac{t}{2}) \quad &\cos(\frac{t}{2})
        \end{pmatrix},
    \end{equation}
    \begin{equation}
        \gamma_2(t) = \begin{pmatrix}
            \cos(\frac{t}{2}) \quad &-\sin(\frac{t}{2}) \\
            \sin(\frac{t}{2}) \quad &\cos(\frac{t}{2})
        \end{pmatrix},
    \end{equation}
    \begin{equation}
        \gamma_3(t) = \begin{pmatrix}
            e^{-\frac{it}{2}} &\quad 0 \\
            0 \quad &e^{\frac{it}{2}}
        \end{pmatrix}.
    \end{equation}
\end{subequations}
A generic state constructed on such cubic graph takes the following form:
\begin{equation}
    D_{\mathrlap{x} \phantom{(j_1)j_1}  j_1}^{(j_1) j_1} (h_1)\,
    D_{\mathrlap{y} \phantom{(j_2)j_2}  j_2}^{(j_2) j_2} (h_2)\,
    D_{\mathrlap{z} \phantom{(j_3)j_3}  j_3}^{(j_3) j_3} (h_3).
    \label{cubic state in Bianchi IX}
\end{equation}
Here, these holonomies intersect each other at both the north and the south poles of the sphere. Consequently, the primary distinction with respect to the Bianchi I case lies in the presence of two vertices. 
We now analyze whether this class of states satisfies the quantum constraints within this framework. The action of the diagonal constraint $\hat{\pazocal{X}}^2$ is identical to that of the preceding case. Therefore, this constraint vanishes automatically on the state introduced in Eq.~\eqref{cubic state in Bianchi IX}:
\begin{equation}
    \hat{\pazocal{X}}^2 \,
    \left[D_{\mathrlap{x} \phantom{(j_1)j_1}  j_1}^{(j_1) j_1} (h_1)\,
    D_{\mathrlap{y} \phantom{(j_2)j_2}  j_2}^{(j_2) j_2} (h_2)\,
    D_{\mathrlap{z} \phantom{(j_3)j_3}  j_3}^{(j_3) j_3} (h_3) \right] = 0.
\end{equation}
Regarding the divergence constraint, it is worth noting that the vanishing result established in Eq.~\eqref{eq. vanishing of the regularized term of div constraint} remains valid. Indeed, the explicit evaluation previously performed is fundamentally independent of the total number of vertices. Specifically, once a particular vertex is selected, one can evaluate the local smearing function at that point and proceed with the identical derivation detailed above, irrespective of the other vertex. Consequently, the regularized operator annihilates the state in the Bianchi IX framework as well:
\begin{equation}
    \begin{split}
        \hat{R}_{i} & \Bigl[ 
        D_{\mathrlap{x} \phantom{(j_1)j_1} j_1}^{(j_1) j_1} (h_1) 
        D_{\mathrlap{y} \phantom{(j_2)j_2} j_2}^{(j_2) j_2} (h_2) 
        D_{\mathrlap{z} \phantom{(j_3)j_3} j_3}^{(j_3) j_3} (h_3) 
        \Bigr] = 0.
    \end{split}
\end{equation}
Regarding the Gauss constraint appearing in the full expression of the divergence constraint, its resolution requires a gauge-invariant state. Such a state is obtained as a specific linear combination of basis states, as proposed, for instance, in \cite{Alesci_Cianfrani_2013a} by using the so-called reduced intertwiners. Once this specific combination is constructed, both the diagonal and the divergence constraints are fully solved, yielding a quantum state that satisfies the quantum counterpart of the classical gauge-fixing conditions standard in LQC.

\section{Conclusion}
In this work, we have investigated the possibility of recovering the standard cosmological sector of Loop Quantum Cosmology directly within the framework of Loop Quantum Gravity by implementing, at the quantum level, the gauge-fixing conditions that characterize the usual minisuperspace description. The starting point is the formulation of homogeneous Ashtekar--Barbero variables that preserves the local \(SU(2)\) gauge freedom of the full theory. Within this framework, homogeneity and diagonality are not imposed a priori at the level of the phase space variables, but are instead understood as gauge-fixing conditions. In particular, we have considered the divergence-free constraint, which selects the homogeneous gauge, together with the diagonal constraint, which restricts the triad to the diagonal sector. At the classical level, the simultaneous imposition of these conditions reproduces the reduced phase space underlying standard Loop Quantum Cosmology.

We first analyzed the consistency of these conditions with the constraint structure of the theory. The Poisson-bracket analysis shows that the divergence-free and diagonal constraints constitute a consistent set of gauge-fixing conditions for the internal \(SU(2)\) and spatial diffeomorphism symmetries, respectively. In particular, the two gauge-fixing conditions commute identically with each other, so that their order of implementation is irrelevant. The corresponding Faddeev--Popov-type matrix has a non-vanishing determinant on the gauge-fixed constraint surface, confirming that the selected conditions provide a consistent gauge fixing of the relevant first-class constraints.

We then constructed quantum operators corresponding to both gauge-fixing conditions using the standard holonomy--flux representation of Loop Quantum Gravity. A relevant feature of our construction is that the regularization is formulated independently of the specific cosmological model under consideration. The diagonal constraint is expressed in terms of flux operators associated with surfaces adapted to the fiducial directions, while the divergence-free condition is regularized through infinitesimal holonomies and fluxes. In this way, both classical gauge-fixing conditions admit well-defined quantum counterparts within the kinematical Hilbert space of LQG.

We have subsequently shown that the quantum gauge-fixing conditions select a distinguished class of homogeneous spin-network states. We constructed homogeneous states on a cubical graph for both Bianchi I and Bianchi IX models, using large spin labels and maximal magnetic quantum numbers in bases adapted to the three fiducial directions. These are precisely the type of states employed in Quantum-Reduced Loop Gravity. The diagonal constraint annihilates these states exactly, while the regularized divergence-free constraint also vanishes. In the latter case, the cancellation follows from the interplay between the \(j=1\) adjoint representation carried by the infinitesimal holonomy and the projectors associated with the different internal bases.

An important consequence of these results concerns the relation between the \(SU(2)\) structure of Loop Quantum Gravity and the Abelian structure characteristic of Loop Quantum Cosmology. The relation with Quantum-Reduced Loop Gravity is particularly significant. The states selected by the gauge-fixing conditions are built from \(SU(2)\) representation matrices, but in the large-spin regime their relevant matrix elements are characterized by maximal magnetic quantum numbers along the three fiducial directions. The resulting structure effectively separates into three commuting \(U(1)\) sectors, reproducing the characteristic \(U(1)^3\) structure encountered in LQC. This suggests that the Abelianization characteristic of the cosmological theory need not be introduced as an independent assumption, but can instead emerge from the quantum implementation of the appropriate gauge-fixing conditions. In this sense, the present construction provides a direct quantum-level link between the homogeneous \(SU(2)\)-covariant formulation of Loop Quantum Gravity and the reduced states used in LQC. The result therefore gives a concrete realization of the idea that the minisuperspace description can be understood as a gauge-fixed sector of a theory that retains the full internal gauge structure prior to gauge fixing.

\section*{Acknowledgment}
The authors would like to express their gratitude to Francesco Cianfrani for his valuable feedback and for the fruitful discussions during the preparation of this work. M.B. acknowledges support by grants ID\# 63132 and ID\# 63683 from the John Templeton Foundation, managed by the Center for SpaceTime and the Quantum.

\bibliographystyle{unsrt}
\bibliography{bibliografia.bib}

\begin{thebibliography}{10}

\bibitem{Thiemann_2007}
Thomas Thiemann.
\newblock {\em Modern Canonical Quantum General Relativity}.
\newblock Cambridge Monographs on Mathematical Physics. Cambridge University Press, Cambridge, 2007.

\bibitem{Rovelli_Vidotto_2014}
Carlo Rovelli and Francesca Vidotto.
\newblock {\em Covariant Loop Quantum Gravity: An Elementary Introduction to Quantum Gravity and Spinfoam Theory}.
\newblock Cambridge University Press, Cambridge, 2014.

\bibitem{Ashtekar_Singh_2011}
Abhay Ashtekar and Parampreet Singh.
\newblock Loop quantum cosmology: a status report.
\newblock {\em Classical and Quantum Gravity}, 28(2121):213001, September 2011.

\bibitem{Banerjee_Calcagni_Martín-Benito_2012}
Kinjal Banerjee, Gianluca Calcagni, and Mercedes Martín-Benito.
\newblock {Introduction to Loop Quantum Cosmology}.
\newblock {\em SIGMA. Symmetry, Integrability and Geometry: Methods and Applications}, 8:016, March 2012.

\bibitem{Bojowald_2000b}
Martin Bojowald.
\newblock {Loop Quantum Cosmology I: Kinematics}.
\newblock {\em Classical and Quantum Gravity}, 17(66):1489–1508, March 2000.

\bibitem{Bojowald_2000a}
Martin Bojowald.
\newblock {Loop quantum cosmology: II. Volume operators}.
\newblock {\em Classical and Quantum Gravity}, 17(66):1509, March 2000.

\bibitem{Bojowald_2001}
Martin Bojowald.
\newblock {Loop quantum cosmology: III. Wheeler-DeWitt operators}.
\newblock {\em Classical and Quantum Gravity}, 18(66):1055, March 2001.

\bibitem{Bojowald_2002}
Martin Bojowald.
\newblock {Isotropic Loop Quantum Cosmology}.
\newblock {\em Classical and Quantum Gravity}, 19(1010):2717–2741, May 2002.

\bibitem{Bojowald_2003}
Martin Bojowald.
\newblock {Homogeneous Loop Quantum Cosmology}.
\newblock {\em Classical and Quantum Gravity}, 20(1313):2595–2615, July 2003.

\bibitem{Bojowald_2020}
Martin Bojowald.
\newblock Critical evaluation of common claims in loop quantum cosmology.
\newblock {\em Universe}, 6(33):36, February 2020.

\bibitem{Ashtekar_Bojowald_Lewandowski_2003}
Abhay Ashtekar, Martin Bojowald, and Jerzy Lewandowski.
\newblock Mathematical structure of loop quantum cosmology.
\newblock {\em Advances in Theoretical and Mathematical Physics}, 7(22):233–268, March 2003.

\bibitem{Brunnemann_Fleischhack_2012}
Johannes Brunnemann and Christian Fleischhack.
\newblock {Non-Almost Periodicity of Parallel Transports for Homogeneous Connections}.
\newblock {\em Mathematical Physics, Analysis and Geometry}, 15(4):299–315, December 2012.

\bibitem{Fleischhack_2018}
Christian Fleischhack.
\newblock {Loop Quantization and Symmetry: Configuration Spaces}.
\newblock {\em Communications in Mathematical Physics}, 360(2):481–521, June 2018.

\bibitem{Bruno_Montani_2023a}
Matteo Bruno and Giovanni Montani.
\newblock {Loop quantum cosmology of nondiagonal Bianchi models}.
\newblock {\em Physical Review D}, 107(1212):126013, June 2023.

\bibitem{Bruno_Montani_2023b}
Matteo Bruno and Giovanni Montani.
\newblock Is the diagonal case a general picture for loop quantum cosmology?
\newblock {\em Physical Review D}, 108(44):046003, August 2023.

\bibitem{Bruno_2025a}
Matteo Bruno.
\newblock {Fiber bundle structure in Ashtekar-Barbero-Immirzi formulation of General Relativity}.
\newblock {\em Journal of Geometry and Physics}, 214:105537, August 2025.

\bibitem{Bruno_2025b}
Matteo Bruno.
\newblock Cosmology in loop quantum gravity: symmetry reduction preserving gauge degrees of freedom.
\newblock {\em Classical and Quantum Gravity}, 42(18):185009, September 2025.

\bibitem{Alesci_Cianfrani_2013a}
Emanuele Alesci and Francesco Cianfrani.
\newblock {Quantum-reduced loop gravity: Cosmology}.
\newblock {\em Physical Review D}, 87(88):083521, April 2013.

\bibitem{Alesci_Cianfrani_2013b}
Emanuele Alesci and Francesco Cianfrani.
\newblock {A new perspective on cosmology in Loop Quantum Gravity}.
\newblock {\em EPL (Europhysics Letters)}, 104(11):10001, October 2013.

\bibitem{Alesci_Cianfrani_Rovelli_2013}
Emanuele Alesci, Francesco Cianfrani, and Carlo Rovelli.
\newblock {Quantum-reduced loop gravity: Relation with the full theory}.
\newblock {\em Physical Review D}, 88(1010):104001, November 2013.

\bibitem{Cianfrani_Marchini_Montani_2012}
Francesco Cianfrani, Andrea Marchini, and Giovanni Montani.
\newblock {The picture of the Bianchi I model via gauge fixing in Loop Quantum Gravity}.
\newblock {\em EPL (Europhysics Letters)}, 99(11):10003, July 2012.

\bibitem{Cianfrani_Montani_2012a}
Francesco Cianfrani and Giovanni Montani.
\newblock {Implications of the gauge-fixing in Loop Quantum Cosmology}.
\newblock {\em Physical Review D}, 85(22):024027, January 2012.

\bibitem{Cianfrani_Montani_2012b}
Francesco Cianfrani and Giovanni Montani.
\newblock {A critical analysis of the cosmological implementation of Loop Quantum Gravity}.
\newblock {\em Modern Physics Letters A}, 27(0707):1250032, March 2012.

\bibitem{Ashtekar_1986}
Abhay Ashtekar.
\newblock {New Variables for Classical and Quantum Gravity}.
\newblock {\em Physical Review Letters}, 57(1818):2244–2247, November 1986.

\bibitem{Ashtekar_1987}
Abhay Ashtekar.
\newblock {New Hamiltonian formulation of general relativity}.
\newblock {\em Physical Review D}, 36(66):1587–1602, September 1987.

\bibitem{Barbero-G._1995}
J.~Fernando Barbero~G.
\newblock {Real Ashtekar variables for Lorentzian signature space-times}.
\newblock {\em Physical Review D}, 51(1010):5507–5510, May 1995.

\bibitem{Immirzi_1997}
Giorgio Immirzi.
\newblock Real and complex connections for canonical gravity.
\newblock {\em Classical and Quantum Gravity}, 14(1010):L177, October 1997.

\bibitem{Ashtekar_Lewandowski_1995}
Abhay Ashtekar and Jerzy Lewandowski.
\newblock Projective techniques and functional integration for gauge theories.
\newblock {\em Journal of Mathematical Physics}, 36(5):2170–2191, May 1995.

\bibitem{Ashtekar_Wilson-Ewing_2009}
Abhay Ashtekar and Edward Wilson-Ewing.
\newblock {Loop quantum cosmology of Bianchi type I models}.
\newblock {\em Physical Review D}, 79(88):083535, April 2009.

\bibitem{Bojowald_2013}
Martin Bojowald.
\newblock {Mathematical Structure of Loop Quantum Cosmology: Homogeneous Models}.
\newblock {\em SIGMA. Symmetry, Integrability and Geometry: Methods and Applications}, 9:082, December 2013.

\bibitem{Henneaux:1992ig}
M.~Henneaux and C.~Teitelboim.
\newblock {\em {Quantization of gauge systems}}.
\newblock Princeton University press, 1992.

\bibitem{doCarmo1992}
Manfredo~P. do~Carmo.
\newblock {\em Riemannian Geometry}.
\newblock Birkhäuser, Boston, 1992.

\bibitem{Carroll:2004st}
Sean~M. Carroll.
\newblock {\em {Spacetime and Geometry}: {An Introduction to General Relativity}}.
\newblock Cambridge University Press, 7 2019.

\bibitem{Bilski_Alesci_Cianfrani_Donà_Marcianò_2017}
Jakub Bilski, Emanuele Alesci, Francesco Cianfrani, Pietro Donà, and Antonino Marcianò.
\newblock {Quantum reduced loop gravity: Extension to gauge vector field}.
\newblock {\em Physical Review D}, 95(10):104048, May 2017.

\bibitem{Makinen_2020}
Ilkka M{\"a}kinen.
\newblock {Operators of quantum-reduced loop gravity from the perspective of full loop quantum gravity}.
\newblock {\em Phys. Rev. D}, 102(10):106010, 2020.

\bibitem{Wilson-Ewing_2010}
Edward Wilson-Ewing.
\newblock {Loop quantum cosmology of Bianchi type IX models}.
\newblock {\em Phys. Rev. D}, 82:043508, 2010.

\end{thebibliography}
\end{document}